\documentclass[aps,pra,final,twocolumn,showpacs,superscriptaddress]{revtex4}
\usepackage{graphicx}
\usepackage{amsmath}

\newcommand{\Ai}{{\rm{Ai}}}
\newcommand{\rmd}{{\rm d}}
\newcommand{\rme}{{\rm e}}
\newcommand{\rmi}{{\rm i}}
\newcommand\encircle[1]{\ensuremath{\mathbin{\settowidth{\dimen7}{\mbox{$\bigcirc$}}
\makebox[0pt][l]{$\bigcirc$}\makebox[\dimen7]{#1}}}}

\begin{document}

\title{Use of Lambert's Theorem for the $n$--Dimensional Coulomb Problem}
\author{Vassiliki Kanellopoulos}
\altaffiliation{Currently Marie Curie Fellow at CERN on the PARTNER ITN, PITN-GA-2008-215840} 
\affiliation{Physik Department T30, Technische Universit{\"a}t M{\"u}nchen James-Franck-Str.,  85747 Garching, Germany.}
\affiliation{Laboratory of Atomic and Solid State Physics, Cornell
University, Ithaca, NY 14853, USA} 
\author{Manfred Kleber}
\affiliation{Physik Department T30, Technische Universit{\"a}t M{\"u}nchen James-Franck-Str.,  85747 Garching, Germany.}
\author{Tobias Kramer}
\affiliation{Institute for Theoretical Physics, Universit{\"a}t Regensburg, 93040 Regensburg, Germany.}
\affiliation{Department of Physics, Harvard University, 17 Oxford Street, Cambridge, MA 02138, USA.}

\begin{abstract}
We present the analytical solution in closed form for the semiclassical limit of
the quantum mechanical Coulomb Green function in position space in
$n$ dimensions. We utilize a projection method which has its roots
in Lambert's theorem and which allows us to treat the system as an
essentially one dimensional problem. The semiclassical result
assumes a simple analytical form and is well suited for a numerical
evaluation. The method can also be extended to classically forbidden
space regions. Already for moderately large principal quantum
numbers $\nu \geq 5$, the semiclassical Green function is found to
be an excellent approximation to the quantum mechanical Green
function.
\end{abstract}


\date{April 21, 2009}

\maketitle

\section{Introduction}

The laws of planetary motion remained for a long time a
mind--puzzling challenge. It was Johannes Kepler who published 400
years ago his book {\it Astronomia Nova} which contained his famous
first two laws on planetary motion. Kepler's conclusion that all
planets move in elliptical orbits with the Sun in one focus was
based on his ingenious evaluation of very accurate observations of
the path of the planet Mars by the astronomer Tycho Brahe, the last
of many important astronomers who made their observations without
the help of a telescope. As is well known, the mathematical
construction scheme for expressing the motion of bodies in a
gravitational ($1/r$) potential in mathematical terms goes back to
the days of Newton's {\it Principia Mathematica}, first published in
1687. This work unifies Galileo's ideas about motion in a
gravitational field and Kepler's laws on planetary motion. \\
In the $18^{th}$ century it was still a major problem to follow the
motion of a planet along its elliptical path or, more general, along
a curved trajectory. For the $1/r$--potential this difficulty was
solved by the Swiss Mathematician and Physicist Johann Heinrich
Lambert who proved geometrically that the transfer time along a
planetary orbit connecting two position vectors $\mathbf{r}$ and
$\mathbf{r^{\prime}}$ depends only on the two combinations
$\alpha_+$ and $\alpha_-$,
\begin{equation}\label{eq:alfaplusminus}
\alpha_+ = r + r^{\prime} + s \;\; \mbox{and} \;\; \alpha_- = r +
r^{\prime} - s \;,
\end{equation}
where $s$ is the distance  between $\mathbf{r}$ and
$\mathbf{r^{\prime}}$. The position vectors are meant relative to
the force center (in Lambert's case the Sun). The additional
dependence of the travel time on $E$ will be discussed later.
Equation (\ref{eq:alfaplusminus}) is a peculiarity of the
$1/r$--potential. The fact that the transfer time depends only on
$\alpha_+$ and
$\alpha_-$ is called Lambert's theorem. \\
The agreement between the calculated and observed positions of the
planets was historically the most important success of classical
physics. With the advent of quantum mechanics, the Kepler problem
was replaced by the Coulomb problem for the hydrogen atom. Feynman's
path integral method revealed the close connection between classical
and quantum mechanics. The fixed--energy propagator for the Coulomb
problem is known analytically both in configuration and momentum
space
\cite{Hostler1964a,Schwinger1964a,Gutzwiller1967a,Gutzwiller1990a,
Schulman1981a,Kleinert1990a,Grosche1998a,Kelsey1976a,Dittrich1999a}. However
the corresponding semiclassical approximation has not been given
before in closed analytic form because of the appearance of a rather
complicated prefactor, the so called Van Vleck--Pauli--Morette
determinant. The semiclassical approach to the Coulomb problem in
$n>1$ dimensions the determinant has been calculated so far only
numerically \cite{Granger2001a}. Based on Lambert's theorem, we will
derive a simple and useful analytic expression for the Van
Vleck--Pauli--Morette determinant in $n$ spatial dimensions. The result
will put us in the position to derive a two-line expression for the 
semiclassical Green function.

\section{Lambert's Theorem for the Reduced Action}\label{action}

\begin{figure}[t]
\begin{center}
\includegraphics[width=0.9\columnwidth]{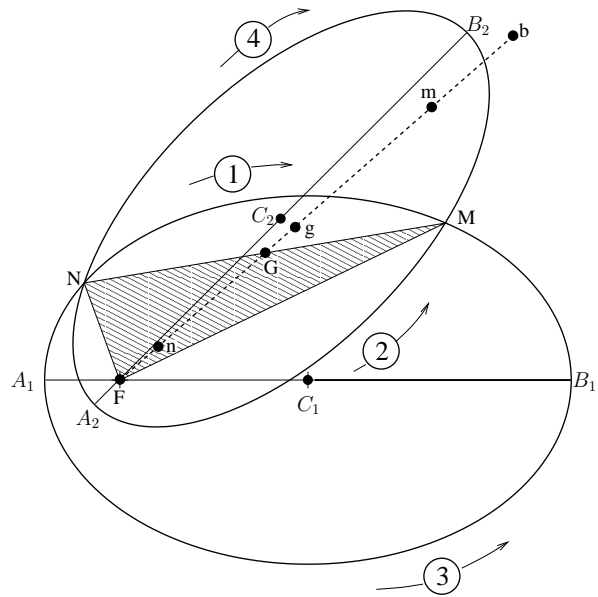}
\end{center}
\caption{Lambert's projection of elliptic motion to collinear
motion. Shown are two ellipses with the same lengths of the
semimajor axes $\frac{1}{2}A_1 B_1=\frac{1}{2}A_2 B_2$ and a common
focus located at $F$. The centers of the two ellipses are denoted by
$C_1$ and $C_2$. Lambert's lemma~24 allows to relate the motion
from $N$ to $M$ on both ellipses to a common collinear motion on the
degenerate linear ellipse $Fb$,  where the points $n$ and $m$ are
chosen such that the time of flight (TOF) along $nm$ equals the TOF
along the elliptical arc $NM$ on the first ellipse. On the second
ellipse the TOF along the arc $NB_2M$ equals the TOF along $nbm$.
The points $n$ and $m$ are found by marking the point $G$ halfway
between $N$ and $M$. Then the major axis $Fb=A_1 B_1=A_2 B_2$ of the
linear ellipse is drawn starting at $F$ and running through $G$. On
this line the point $g$ is placed at the distance
$Fg=\frac{1}{2}(FN+FM)$. Finally $n$ and $m$ are given by the
intersection points of a circle around $g$ with radius $GN=GM$. This
construction shows that the sum of the lengths of the shaded
triangle $\alpha_{\pm}=FN + FM \pm NM$ is equal to $\alpha_{\pm}=Fn
+ Fm \pm nm$. The ficticious collinear motion goes back to Lambert and
can be picturized as the
limit of an elliptic motion with extremely small semiminor axis $b$.
The eccentricity approaches one from below in such a way that the
moving particle turns around at $F$ with very high but of course
still non-relativistic velocity.}
\label{fig:lambert}
\end{figure}

It is a simple exercise in classical mechanics to analyze the
relative motion for the Kepler or Coulomb Hamiltonian
\begin{equation}\label{eq:hamiltonian}
H = \frac{\mathbf{p}^2}{2\mu} - \frac{K_c}{r} \;,
\end{equation}
where $\mu$ is the reduced mass and $K_c$ the strength of
the attractive $1/r$ potential. The corresponding motion in time is
given by
\begin{equation}\label{eq:landau}
t - t' =\sqrt{\frac{\mu a}{K_c}}\int_{r'}^{r}
\frac{\tilde{r}}{\sqrt{2a \tilde{r} - \tilde{r}^2 -a\Lambda^2/(\mu
K_c)}}\,\rmd \tilde{r} \;.
\end{equation}
Here $\Lambda = \mu r^2 \dot \phi$ is the angular momentum
about the center of force for elliptic motion with semimajor axis $a
= K_c/(2|E|)$, for $E<0$.\\
An important element for the transition from classical
mechanics to quantum mechanics is the reduced action $W$, also
called action integral $S$. In order to avoid confusion, we reserve
$S$ here for Hamilton's principal function (see below). Within the
time--independent Hamilton--Jacobi theory the reduced action $W$ is
given by
\begin{equation}\label{eq:action}
W(\mathbf{r},\mathbf{r}';E)=
\int_{\mathbf{r}'}^{\mathbf{r}}\mathbf{p}(\tilde{\mathbf{r}})\cdot\rmd\tilde{\mathbf{r}} \;.
\end{equation}
For elliptic motion in the $x$--$y$--plane an explicit
evaluation of Eq.~(\ref{eq:action}) is easily achieved by
introducing for example Cartesian coordinates with the origin at the
center of the ellipse
\begin{equation}\label{eq:Cartesian}
x = a \cos \xi \,,\,\,\, \,\, y= b \sin \xi \,,
\end{equation}
where $b= a \sqrt{1-\epsilon^2}$ is the semiminor axis of
the Kepler ellipse with eccentricity $\epsilon $. If we substitute
Eq.~(\ref{eq:Cartesian}) into Eq.~(\ref{eq:action}) and use
\begin{equation}\label{eq:time}
t - t' = \sqrt{\frac{\mu a^3}{K_c}} (\xi - \epsilon \sin \xi - \xi'
+ \epsilon \sin \xi')
\end{equation}
for the transfer time between two points on the ellipse, we obtain
\begin{equation}\label{eq:action0}
W(\mathbf{r},\mathbf{r}';E)= \sqrt{\mu a K_c}\,\, ( \xi + \epsilon \sin \xi -
\xi' - \epsilon \sin \xi' ).
\end{equation}
$W(\mathbf{r},\mathbf{r}';E) $ is a function of $E$ and of the
initial and final coordinates $\mathbf{r}$ and $\mathbf{r}'$ of the planet.
Therefore other dynamical quantities, like the orbital angular
momentum $\Lambda$  must be eliminated. Hence we have to get rid of
$\epsilon = \sqrt{1- 2|E|\Lambda^2/(\mu K_c^2)}$ in
Eq.~(\ref{eq:action0}). A few algebraic manipulations (see App.~\ref{sec:transformation}) lead to
\begin{equation}\label{eq:action2} W(\mathbf{r},\mathbf{r}';E) =
\sqrt{\mu a K_c}\,\, ( \gamma + \sin \gamma - \delta - \sin \delta
)\,
\end{equation}
with
\begin{equation}\label{eq:definition}
\sin^2\frac{\gamma}{2} =  \frac{r + r' + s}{4a} \quad {\rm and}
\quad \sin^2\frac{\delta}{2} = \frac{r + r' - s}{4a} \,.
\end{equation}
In the last equation and in what follows $r$ and $r'$ are
the distances from the focus of the ellipse (i.~e. the center of
force) to two arbitrary points on the elliptical orbit. As before
$s$ stands for the distance between $r$ and $r'$. The situation is
depicted in Fig.~\ref{fig:lambert}.
It was Lambert \cite{Lambert1761a} who succeeded to map elliptical motion to
collinear motion. He also proved (\cite{Lambert1761a}, Lemma 24)
that for fixed energy $E<0$, the classically allowed elliptic motion
from a given initial point $N$ to a final point $M$ can generally
occur on two different ellipses unless we have circular motion
(compare \cite{Gutzwiller1990a}, p.~27). From the last two equations
it becomes obvious that the reduced action $W$ is a function of
$\alpha_+ = r +r' +s$ and $\alpha_- = r + r' -s$. Note that the
energy dependence of $W$ enters
through the semimajor axis $a = K_c/(2|E|)$. \\
Another piece of information is Hamilton's principal function
$S(\mathbf{r},\mathbf{r}',\tau)$ which follows from the well--known Legendre
transformation
\begin{equation}\label{eq:legendre}
S(\mathbf{r},\mathbf{r}',\tau) = W(\mathbf{r},\mathbf{r}',E) - E\tau\,.
\end{equation}
The travel time  $\tau = t-t'$ from $\mathbf{r'}$ to $\mathbf{r}$
can be calculated from
\begin{equation}\label{eq:time1}
\tau = \frac{\partial W}{\partial E}\,,
\end{equation}
or directly from Eq.~(\ref{eq:time}) by using the method of 
App.~\ref{sec:transformation}. The result is
\begin{equation}\label{eq:time2}
\tau = t-t'  = \sqrt{\frac{\mu a^3}{K_c}}(\gamma - \sin \gamma - \delta + \sin \delta) \, .
\end{equation}
Equation (\ref{eq:time2}) is Lambert's theorem (\cite{Lambert1761a},
p.~102) for the travel time between $N$ and $M$. In our case it is
more important to point out that Lambert's theorem is not only valid
for the travel time but also for the reduced action $W$ and,
although not of importance here, for Hamilton's principal function
$S$. With these results in mind, we are now in a position to
calculate the semiclassical Green function.

\section{Coulomb Trajectories and Lambert's Theorem}\label{sec:uselambert}

In quantum physics, the Kepler problem becomes the Coulomb
problem. The connection between classical and quantum mechanics is
conveniently established through the introduction of the quantum
mechanical Green function, also called propagator. The Green
function is the mathematical vehicle that allows a particle to go
from an initial configuration to a final one. In configuration space
it represents the transition amplitude to travel from $\mathbf{r'}$
to $\mathbf{r}$. Each classical trajectory in Fig.~\ref{fig:lambert} has sharp
energy and travel time. In quantum mechanics the travel may occur
either with fixed energy or in a given time. Travel with fixed
energy is characterized by the nonrelativistic energy Coulomb Green
function which was obtained by Hostler \cite{Hostler1963a} in
configuration space in closed form, starting from a partial-wave
expansion. The Coulomb Green function in momentum space
was derived soon after by Schwinger \cite{Schwinger1964a}.\\
Feynman's path integral method is a natural way to
calculate transition amplitudes. For classically allowed transitions
one has to identify all classically allowed trajectories, assign
each of them with a phase and an amplitude and sum up their
contribution. This procedure yields as an approximation the
semiclassical amplitude. In a quantum mechanically exact calculation
of the propagator one would have to sum up all paths, including the
classically forbidden ones. Semiclassical methods work usually very
well because classical trajectories carry the main information
needed to calculate the transition amplitude from $\mathbf{r'}$ to
$\mathbf{r}$ \cite{Brack1997a}. In particular, semiclassical methods are accurate and
useful when large angular momenta are involved. Typical problems
with high--angular wave packets require in an exact quantal
calculation a non--trivial summation over many partial waves of the
Green function. This problem is avoided in the semiclassical
treatment presented here where no summation over partial waves is necessary.
\begin{figure}[t]
\begin{center}
\includegraphics[width=0.9\columnwidth]{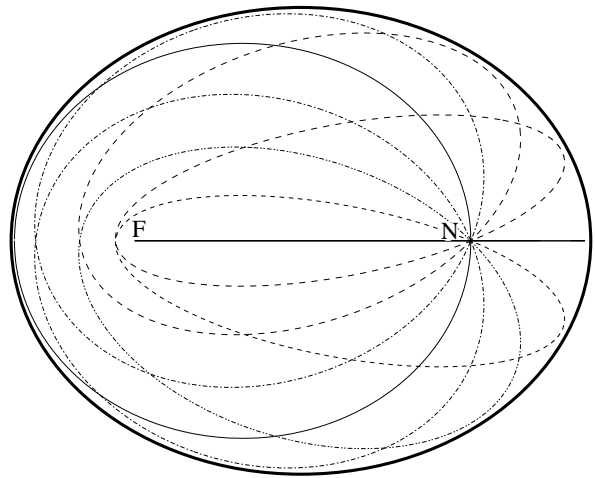}
\end{center}
\caption{The caustic (thick solid line) for the Kepler problem is an
ellipse. Classically allowed orbits with the same energy, the same
center of force $F$ and a common starting point $N$ have to lie
inside this critical ellipse.\label{fig:kaustik1}}
\end{figure}
For $E<0$ a particle will move on an ellipse in a plane
with the center of force in one focus. The binding energy fixes the
length of the semimajor axis while the semiminor axis will also
depend on the angular momentum. As shown in
Fig.~\ref{fig:kaustik1}, the classically allowed trajectories are
confined to the volume in position space defined by the equations of
motions for a given initial absolute value of the velocity. In two
dimensions, one has a critical ellipse that leads to a finite
classical motion.\\
From the definition (1) of $\alpha_+$ and $\alpha_-$ it
follows that $\alpha_+/2$ and $\alpha_-/2$ are the distances of $m$
and $n$ from point $F$ (see Fig.~\ref{fig:lambert}). Hence we identify
the distances $\alpha_\pm$ as path coordinates of $n$ and $m$ along
the straight line $\overline{Fnm}$ with $F$ as origin. Energy
conservation $H=E$ in Eq.~(\ref{eq:hamiltonian}) yields the
velocities
\begin{equation}\label{eq:vel1}
v_{\pm}=\sqrt{\frac{2 |E|}{\mu}}\sqrt{\frac{4
a-\alpha_{\pm}}{\alpha{\pm}}}
\end{equation}
in $m$ and $n$. By making use of the coordinates $\alpha/2$ and
velocities $v(\alpha)$ we obtain the reduced action for travelling
from $n$ to $m$:
\begin{equation}\label{eq:action1}
W(\mathbf{r},\mathbf{r}';E)=  W_+(\alpha_+,E) - W_-(\alpha_-,E)\,,
\end{equation}
where
\begin{equation}\label{eq:wirkungen}
\begin{split}
W_\pm&=\mu \int_0^{\alpha_\pm/2}
v_+\,\rmd (\tilde{\alpha}_\pm/2)\\
&=\mu
\int_0^{\alpha_\pm/2}\sqrt{\frac{K_c}{\mu a}} \frac{\sqrt{a
\tilde{\alpha}_\pm-{(\tilde{\alpha}_\pm/2)}^2}}{\tilde{\alpha}_\pm/2}\,\rmd
(\tilde{\alpha}_\pm/2)\;
\end{split}
\end{equation}
can be cast in closed form
\begin{equation}\label{eq:wirkungen1}
W_\pm = \sqrt{\frac{K_c \mu}{a}}\left(\frac{1}{2}\sqrt{(4
a-\alpha_\pm)\alpha_\pm}+ 2
a\arctan{\sqrt{\frac{\alpha_\pm}{4a-\alpha_\pm}}}\right).
\end{equation}
The last three equations are consistent with Eqs.~(\ref{eq:action2})
and (\ref{eq:definition}) and confirm the essentially
one--dimensional character of the reduced action for the Kepler and
Coulomb problem.

\section{Semiclassical Energy Green Function}\label{sec:SC}

\begin{figure}[t]
\begin{center}
\includegraphics[width=0.8\columnwidth]{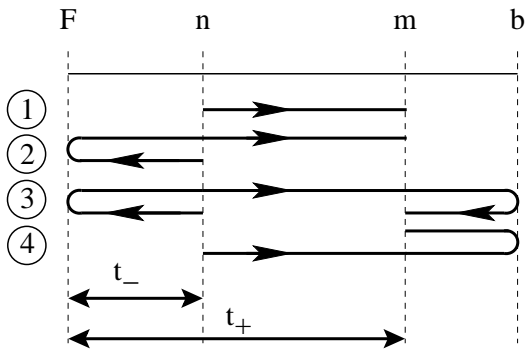}
\end{center}
\caption{
The four elementary paths from $n$ to $m$ according to
Lambert's mapping theorem. The travel time for each path can be
expressed by $t_-$ , $t_+$ and the time for a round trip.
}
\label{fig:fourpaths}
\end{figure}

The $n$--dimensional energy Green function is a solution of the
inhomogeneous stationary Schr\"odinger equation
\begin{equation}
\left[E-H \right]G^{(n)}(\mathbf{r},\mathbf{r}';E)=\delta^{(n)}(\mathbf{r}-\mathbf{r}').
\end{equation}
with $\delta^{(n)}$ being a delta--function point source in $n$
dimensions. Different boundary conditions on $G^{(n)}$ are possible.
For scattering problems, outgoing--wave boundary conditions are
usually appropriate. For standing waves and for bound--state
problems $G^{(n)}$ is real. $G^{(n)}(\mathbf{r},\mathbf{r}';E)$ characterizes the
probability-amplitude for traveling from $\mathbf{r}$ to $\mathbf{r}'$ with a
given energy $E$. For the $n$-dimensional Coulomb problem the
Hamiltonian is given by
\begin{equation}
H=-\frac{\hbar^2}{2 m}\mathbf{\Delta}+\frac{K_c}{r}\,,
\end{equation}
where $\mathbf{\Delta}$ is the Laplace operator and $r$ the distance
from the force center in $n$
dimensions.\\
The semiclassical limit of the energy Green function is
given by
\cite{Gutzwiller1990a,Schulman1981a,Kleinert1990a,Grosche1998a}
\begin{equation}\label{eq:elG}
\begin{split}
G_\text{sc}^{(n)}(\mathbf{r},\mathbf{r}';E) &=\frac{1}{\rmi \hbar} \sum_{i}
\frac{-1}{(-2\pi\rmi\hbar)^{(n-1)/2}}
\,{\left|\mathbf{D}(W_i(\mathbf{r},\mathbf{r}';E))\right|}^{1/2}\\
&\quad\quad\times
\exp\left[\frac{\rmi}{\hbar}W_{i}(\mathbf{r},\mathbf{r}';E)-\rmi
m_i\frac{\pi}{2} \right]\,,
\end{split}
\end{equation}
where
\begin{equation}\label{eq:VV1}
\mathbf{D}(W(\mathbf{r},\mathbf{r}';E))=\det\left(
\begin{array}{cc}
  \frac{\partial^2 W}{\partial \mathbf{r} \partial \mathbf{r}'}& \frac{\partial^2 W}{\partial \mathbf{r} \partial E}\\
\frac{\partial^2 W}{\partial E \partial \mathbf{r}'}& \frac{\partial^2
W}{\partial E^2}
\end{array}\right)
\end{equation}
is the Van Vleck--Pauli--Morette (VVPM) determinant. In
Eq.~(\ref{eq:elG}) one has to sum over all classical fixed--energy
paths $i$ leading from $\mathbf{r}'$ to $\mathbf{r}$ and having the reduced
action $W_i$. The VVPM--determinant contains derivatives of second
order with respect to $\mathbf{r}$, $\mathbf{r}'$ and $E$. For example,
$\frac{\partial^2 W}{\partial \mathbf{r} \partial \mathbf{r}'}$ is a $n\times
n$ matrix with mixed derivatives with respect to starting (initial)
and ending (final) points $\mathbf{r}'=(x'_1,x'_2,...,x'_n)$ and
$\mathbf{r}=(x_1,x_2,...,x_n)$. \\
Finally, $m_i$ is the Morse index which is the number of conjugate
points along the trajectory from $\mathbf{r}'=(x'_1,x'_2,...,x'_n)$ to
$\mathbf{r}=(x_1,x_2,...,x_n)$. In the next section the indices will be
read off from the analytical result for the VVPM--determinant. We
have seen before that Lambert's theorem allows the Coulomb problem
to be mapped on a 1--D problem.
\begin{table*}[t]
\begin{center}
\begin{tabular}{lllc}
path & action & travel time & Morse index \\\hline
\encircle{1} direct path       & $W_1= W_+ -W_-$  & $T_1 = t_+-t_-$ & 0\\
\encircle{2} reflection at $F$ & $W_2 = W_+ +W_-$ & $T_2= t_++t_-$  & 1 \\
\encircle{3} two reflections   & $W_3 = W_{2\pi} + (W_+ -W_-)$ & $T_3 = T_{2\pi} - (t_+ - t_-)$ & 2 \\
\encircle{4} reflection at the caustic & $W_4 = W_{2\pi}- (W_+ + W_-)$ & $T_4 = T_{2\pi}-(t_+ + t_-)$ & 1
\end{tabular}
\end{center}
\caption{Reduced action and Morse-indices $m_i$ for bounded motion
along the four elementary trajectories in classically
allowed regions for an attractive Coulomb potential in three
dimensions (see also Fig.~\ref{fig:lambert}). The reduced actions are combinations
of $W_+$ and $W_-$ and the Morse indices in $n$ dimensions are obtained in Sec.~\ref{sec:vanvleck}.}
\label{tab:fourpaths}
\end{table*}
Utilizing Lambert's projection theorem, we are now in a
position to find {\it{all}} possible trajectories and, if needed,
all traveling times. 
Figure~\ref{fig:fourpaths} reveals all
elementary possibilities to travel from $n$ to $m$. As in 
Fig.~\ref{fig:lambert} already mentioned, we regard the motion in 1-D as
motion on an ellipse with infinitesimally small semiminor axis $b$.
For such motion we obtain Table~\ref{tab:fourpaths}, where
\begin{equation}\label{eq:roundtrip}
W_{2\pi} = 2\pi\sqrt{\mu a K_c} \quad {\rm and} \quad T_{2\pi}
= 2\pi\sqrt{\frac{\mu a^3}{K_c}}
\end{equation}
denote the action for a closed orbit and the time of circulation on
the same closed orbit respectively. We observe that both quantities
depend only on the semimajor axis $a$, i.e. on the orbital
energy $E$.\\
Traveling from $\mathbf{r}'$ to $\mathbf{r}$ at constant energy is possible
along one of the four elementary paths. However there is an infinite
number of possibilities for traveling due to the addition of an
arbitrary number of loops to each elementary path.

\section{The Van Vleck--Pauli--Morette Determinant}\label{sec:vanvleck}

We will now calculate the amplitude of the Green's function, i.e.
the VVPM determinant $\mathbf{D}(W)$~(Eq.~\ref{eq:VV1}).
It is helpful to realize that the  $n\times n$ sub-determinant
$\left|\frac{\partial^2 W}{\partial \mathbf{r}\partial\mathbf{r}'}\right|$
vanishes (\cite{Gutzwiller1990a}, page~24). Therefore
$\mathbf{D}(W)$ is independent of the matrix element
$\left|\frac{\partial^2 W}{\partial E^2}\right|$. We will replace
this element by~$0$. The VVPM matrix contains mixed second 
derivatives of $W$ with respect to the coordinates $x_i'$ and $x_j$.
In the last section we showed how to express the reduced action for
the elementary four paths as combinations of the two basic actions
$W_+(\alpha_+(\mathbf{r},\mathbf{r}'))$ and $W_-(\alpha_-(\mathbf{r},\mathbf{r}'))$
(Fig.~\ref{fig:fourpaths} and Table~\ref{tab:fourpaths}). By using $\frac{\partial W_\pm}{\partial
(\alpha_\pm/2)}=\mu v_\pm$  together with the chain rule we find for
the off diagonal $(i \neq j)$ elements
\begin{widetext}
\begin{equation}\label{eq:offdiag1}
\frac{\partial^2 W_+}{\partial x_i \partial  x_j'}=\frac{\mu}{4}
\frac{\partial v_+}{\partial (\alpha_+/2)}\left( \frac{x_j'}{r'}-
\frac{x_j-x_j'}{s}\right)\left(
\frac{x_i}{r}+\frac{x_i+x_i'}{s}\right) +\frac{\mu
v_+}{2}\frac{(x_i-x_i')(x_j-x_j')}{s^3}\,,
\end{equation}
\begin{equation}\label{eq:offdiag2}
\frac{\partial^2 W_-}{\partial x_i \partial  x_j'}=\frac{\mu}{4}
\frac{\partial v_-}{\partial (\alpha_-/2)}\left(
\frac{x_j'}{r'}+\frac{x_j-x_j'}{s}\right)\left(
\frac{x_i}{r}-\frac{x_i+x_i'}{s}\right)-\frac{\mu
v_-}{2}\frac{(x_i-x_i')(x_j-x_j')}{s^3}\,,
\end{equation}
where we made use of the fact that according to Eq~(\ref{eq:vel1})
the $v_\pm$ are functions of $\alpha_\pm$ . The diagonal elements
follow in a similar fashion
\begin{equation}\label{eq:ondiag1}
\frac{\partial^2 W_+}{\partial x_j \partial  x_j'}=\frac{\mu}{4}
\frac{\partial v_+}{\partial (\alpha_+/2)}\left(
 \frac{x_j'}{r'}-\frac{x_j-x_j'}{s}\right)\left(
 \frac{x_j}{r}+\frac{x_j+x_j'}{s}\right)-
 \frac{\mu v_+}{2 s}\left(1-\frac{(x_j-x_j')^2}{s^2}\right)\,,
\end{equation}
\begin{equation}\label{eq:ondiag2}
\frac{\partial^2 W_-}{\partial x_j \partial  x_j'}=\frac{\mu}{4}
\frac{\partial v_-}{\partial (\alpha_-/2)}\left( \frac{x_j'}{r'}+
\frac{x_j-x_j'}{s}\right)\left(
\frac{x_j}{r}-\frac{x_j+x_j'}{s}\right)+ \frac{\mu v_-}{2 s}
\left(1-\frac{(x_j-x_j')^2}{s^2}\right)\,.
\end{equation}
\end{widetext}
The mixed derivatives with respect to energy and coordinates are
obtained by utilizing $\frac{\partial W}{\partial E}=t$. We also
have $\frac{\partial W_\pm}{\partial E}=t_\pm$.  Lambert's
projection of the Coulomb problem to a linear, one dimensional
motion implies that
\begin{equation}\label{eq:relation}
\frac{\partial^2 W_\pm}{\partial(\alpha_\pm/2)\partial E} =
\frac{\partial t_\pm}{\partial(\alpha_\pm/2)}=\frac{\rmd t_\pm
}{\rmd (\alpha_\pm/2)}=\frac{1}{v_\pm}.
\end{equation}
We can therefore write
\begin{equation}\label{eq:energie1}
\frac{\partial^2 W_+}{\partial x_{j} \partial E}=\frac{ \partial
t_+}{ \partial (\alpha_+/2)} \frac{ \partial(\alpha_+/2)}{\partial
x_{j}} =\frac{1}{v_+}\left(\frac{x_j}{2 r}+ \frac{x_{j}- x_{j}'}{2
s} \right)\,,
\end{equation}
\begin{equation}\label{eq:energie2}
\frac{\partial^2 W_-}{\partial x_{j} \partial E}=\frac{1}{v_-}
\left(\frac{x_j}{2 r}- \frac{x_{j}- x_{j}'}{2s} \right).
\end{equation}
Finally,
\begin{equation}\label{eq:energie1a}
\frac{\partial^2 W_+}{\partial x_{i}' \partial E}=\frac{ \partial
t_+}{ \partial (\alpha_+/2)}\frac{ \partial(\alpha_+/2)}{\partial
x_{i}'}=\frac{1}{v_+}\left(\frac{x_i'}{2 r'}- \frac{x_{i}- x_{i}'}{2
s} \right)\,,
\end{equation}
\begin{equation}\label{eq:energie2a}
\frac{\partial^2 W_-}{\partial x_{i}' \partial
E}=\frac{1}{v_-}\left(\frac{x_i'}{2 r'}+ \frac{x_{i}- x_{i}'}{2s}
\right)\,.
\end{equation}
In principle it is possible to evaluate the second derivatives for
elliptic motion for all points $N$ and $M$. However
this is a tedious task. Fortunately Lambert's theorem tells us that
elliptical Kepler motion can be mapped on a degenerate ellipse where
motion occurs on a 1--D straight line. We use this mapping and
assume the coordinate $q$ to run along this line from $ q' =
\alpha_-/2$ to $q = \alpha_+/2$, i.e.\ from point $n$ to point $m$.
In $n$ dimensions we have $n-1$ coordinates $x_2, x_3, \ldots
x_n$ that are orthogonal to the trajectory. Along the trajectory we
have $x_i = x_i' = 0$ for $i \geq 2$. If we therefore evaluate
Eqs.~(\ref{eq:offdiag1})-(\ref{eq:energie2a}) for $i,j \geq 2$, we
observe that the right hand sides vanish except for the diagonal
matrix elements
\begin{equation}\label{eq:reduce1}
\mathcal{F}_+  :=  \frac{\partial^2 W_+}{\partial x_j
\partial x_j'}|_{x_j,x_j'=0}=-\frac{\mu v_+}{2 s}\,,\quad j\geq2\,,
\end{equation}
and
\begin{equation}\label{eq:reduce2}
\mathcal{F}_-  := \frac{\partial^2 W_-}{\partial x_j \partial
x_j'}|_{x_j,x_j'=0}=\frac{\mu v_-}{2 s}\,,\quad j\geq2\,.
\end{equation}
Each direction $i \geq 2$ orthogonal to the straight--line
trajectory contributes with the same dimensional factor $\cal{F}$.
Putting everything together we can cast the $(n+1)\times (n+1)$ VVPM
determinant in a simple form:
\begin{equation}\label{eq:matrixvvpm}
\mathbf{D}(W) = \det\left(\begin{array}{ccccc}
\frac{\partial^2 W}{\partial (\alpha_+/2) \partial (\alpha_-/2)} & 0 & \cdots & 0 & \frac{\partial^2 W}{\partial (\alpha_+/2)\partial E}\\
  0     & \mathcal{F}   &       &           &0  \\
\vdots  &       &\ddots &           & \vdots \\
0       &       &           &\mathcal{F}    &0\\
 \frac{\partial^2 W}{\partial (\alpha_-/2) \partial E}  & 0 & \cdots &0          & 0
\end{array}\right).
\end{equation}
From Table~\ref{tab:fourpaths} we infer that the action $W$ needed for the four
elementary paths is always a linear combination of $W_+$ and $W_-$.
The necessary determinants $\mathbf{D}(W_+ \pm W_-)$ are obtained
from Eq.~(\ref{eq:matrixvvpm}) with $\mathcal{F}$ replaced by $\mathcal{F}_+ \pm
\mathcal{F}_-$. Recalling that $W_+$ ($W_-$) is a function
$\alpha_+$ ($\alpha_-$) only, we conclude that
\begin{equation}
\frac{\partial^2 W}{\partial (\alpha_+/2) \partial (\alpha_-/2)} =
0.
\end{equation}
Therefore the  entry on the top left of VVPM--matrix vanishes. The
determinant is now easily calculated via Laplace expansion. The
result is
\begin{equation}\label{eq:result}
\mathbf{D}(W)  =-\frac{\partial^2 W}{\partial (\alpha_+/2)
\partial E}\frac{\partial^2 W}{\partial  (\alpha_-/2)
\partial E}\times \mathcal{F}^{(n-1)}.
\end{equation}
A straightforward evaluation of Eq.~(\ref{eq:result}) yields simple
results for the determinants of the four elementary
paths: \\
\begin{equation}\label{eq:VV}
\begin{split}
\mathbf{D}\encircle{1} 
&= \mathbf{D}(W_+-W_-)=\frac{1}{v_+v_-}\left[-\frac{\mu}{2s}(v_++v_-)\right]^{(n-1)} \\
&= -\mathbf{D} \encircle{3}\,,
\end{split}
\end{equation}
\begin{equation}\label{eq:VVV}
\begin{split}
\mathbf{D}\encircle{2}
&= \mathbf{D}(W_++W_-)=-\frac{1}{v_+v_-}\left
[-\frac{\mu}{2s}(v_+-v_-)\right]^{(n-1)} \\
&= - \mathbf{D} \encircle{4}\,.
\end{split}
\end{equation}
We should point out that Eq.~(\ref{eq:result}) is also valid for
scattering states if action and velocities are adapted to unbounded
motion. \\ 
We now determine the \textit{Morse indices} $m_i$ which are given by the order of
the zeros of the determinants $\mathbf{D}\encircle{i}$ along path
number $i$. Here we restrict ourselves to the three--dimensional
Coulomb problem, $n=3$.  By inspecting Fig.~\ref{fig:fourpaths} we
see that on path~$\encircle{1}$ the velocities are different from zero
because we have assumed that neither point $m$ nor point $n$ is
lying on the caustic $b$. Hence we have $m_1 = 0$. On
path~$\encircle{4}$ the velocity vanishes at the reflection point $b$.
There a pole of first order is generated in the determinant. As a
result we have $m_4 = 1$, independent of $n$. Path~$\encircle{2}$ corresponds to elliptic
motion with infinitesimally small semiminor axis $b$ with the
particle (planet or electron) moving around $F$ with infinite
velocity, $v \rightarrow \infty$. Along this path it therefore
encounters a pole of order $n-2 = 1$ at $F$,  meaning that the
particle picks up the Morse index $m_2= 1$. Obviously we have $m_3
= m_2 + m_4 =2$. Finally by inspecting Eq.~(\ref{eq:VV}) we observe
that a full round trip picks up an additional phase which originates
from closing the loop with $v_+ = v_-$ and $s=0$, giving rise to a
pole of order $n-1 = 2$ in the determinant. In other words, closed
orbits pick the phase $m_{2\pi} = 2(n-1)$.

\section{$E < 0 : $ The Bound--State Green Function}

Having found the amplitudes, reduced actions and the correct phases
we are in a position to evaluate the semiclassical Green function in
analytic form. We showed before that
$G^{(n)}_{\rm{sc}}(\mathbf{r},\mathbf{r'};E)$ consists of the
amplitudes for the four elementary trajectories plus a summation
over all possible loops for each elementary path. Therefore we can
write
\begin{widetext}
\begin{equation}\label{eq:GSC}
\begin{split}
G_{\rm{sc}}^{(n)}(\mathbf{r},\mathbf{r}';E)=&\frac{1}{\rmi \hbar}
\frac{-1}{(-2\pi\rmi\hbar)^{(n-1)/2}} \sum_{i=1}^4 \sum_{j=0}^\infty
\sqrt{|\mathbf{D}(W_i(\mathbf{r},\mathbf{r}';E) +j\,W_{2\pi}(E))|}\\
&\times\exp\left[\frac{\rmi}{\hbar}(W_i(\mathbf{r},\mathbf{r}';E) + j\,W_{2\pi}(E))-\rmi \frac{\pi}{2}(m_i+j\, m_{2\pi})\right]\\
=&G_{\rm elem}(\mathbf{r},\mathbf{r}';E)\times P_{\rm glob}(W_{2 \pi},n).
\end{split}
\end{equation}
\end{widetext}
as a product of the elementary four-path Green function $G_{\rm
elem}(\mathbf{r},\mathbf{r}';E)$ and and a factor  $P_{\rm glob}(W_{2
\pi},n)=\sum_{j=0}^\infty \exp\left[\rmi j
\left(\frac{W_{2\pi}}{\hbar}-\frac{\pi}{2}m_{2\pi}\right)\right] $
which accounts for the loop summation. The factorization is possible
because $\mathbf{D}(W_i(\mathbf{r},\mathbf{r}';E) +j\,W_{2\pi}(E))$ is
independent of $W_{2\pi}(E)$. Each loop adds the same non--negative
phase to the Green function. The summation over the infinite number
of loops can be carried out. We obtain
\begin{equation}\label{eq:GammaSumme1}
\begin{split}
&
\sum_{j=0}^\infty\exp\left[2\pi\rmi j \left(\frac{W_{2\pi}}{2
\pi\hbar}-\frac{m_{2\pi}}{4}\right)\right]
\\\quad&
=\frac{1}{2}+\frac{\rmi}{2}\cot\left[\pi
\left (\frac{W_{2\pi}}{2 \pi\hbar}-\frac{m_{2\pi}}{4}\right
)\right]\,.
\end{split}
\end{equation}
The poles of $P_{\rm glob}(W_{2 \pi},n)$ yield the energy eigenvalues
of the hydrogen atom. Obviously, they are obtained from the poles of
the cotangent given by the non--negative integers,
$\frac{W_{2\pi}}{2\pi\hbar}-\frac{m_{2\pi}}{4}=0,1,2,...$. Using
Eq.~(\ref{eq:roundtrip}) together with $m_{2\pi} = 2(n-1)$, it is now
easy to extract the exact energy eigenvalues for the hydrogen atom
in $n>1$ dimensions \cite{Nieto1979a}
\begin{equation}\label{eq:energyvalues}
E_k=-\frac{\mu
K_c^2}{2\hbar^2}\frac{1}{(k+(n-1)/2)^2}\quad\quad[k=0,1,2,...].
\end{equation}
We should point out that the correct quantization rule for the
action in $n$ dimensions
\begin{equation}\label{eq:quantizationrule}
W_{2\pi}=h (k+\frac{n-1}{2})\quad\quad[k=0,1,2,...]\,
\end{equation}
is an integer multiple of $h$ only for odd values of $n$. \\
The elementary four--path Green function $ G_{\rm elem}$ can be
written in a more compact fashion because paths lying on the same
ellipse have the same amplitude as can be seen from Fig.~\ref{fig:lambert},
and Fig.~\ref{fig:fourpaths}. Their Morse indices are related to each
other through $m_j=(n-1)-m_i$. Therefore we can merge paths
$\encircle{1}$ and $\encircle{3}$ and paths $\encircle{2}$ and $\encircle{4}$
pairwise together. Then the elementary four--path Green function
shows a two-path interference pattern. \\
Putting everything together, we recast the (real) negative energy
Green function (\ref{eq:GSC}) in the form
\begin{equation}\label{eq:Greencomplete}
\begin{split}
G_{\rm{sc}}^{(n)}(\mathbf{r},\mathbf{r}';E) &=
\frac{1}{\hbar(-2\pi\hbar)^{(n-1)/2}}\frac{1}{\sin(k \pi)}\\
&\times\bigg(\sqrt{|\mathbf{D} \encircle{1}|}
\cos\left[\frac{W_1}{\hbar}-\pi\left(\frac{n-1}{4}+k\right)\right]+\\
&\sqrt{|\mathbf{D}\encircle{2}|}\sin\left[\pi\left(\frac{3(n-1)}{4}+k\right)-\frac{W_2}{\hbar}\right]\bigg)\,.
\end{split}
\end{equation}
The bound states (\ref{eq:energyvalues}) at $k = 0,1,2,3, \ldots$
give rise to poles in $G_{\rm{sc}}^{(n)}$. Note that in
Eq.~(\ref{eq:Greencomplete}) $k$ can assume any continuous value $k
\geq 0$. $\mathbf{D}\encircle{1}$ and $\mathbf{D}\encircle{2}$ are
the Van Vleck--Pauli--Morette determinants given before in
Eqs.~(\ref{eq:VV}) and (\ref{eq:VVV}). The actions $W_1=W_+-W_-$ and
$W_2=W_++W_-$ are readily calculated from Eq.~(\ref{eq:wirkungen1}).
Equation (\ref{eq:Greencomplete}) is the main result of the paper.
In the next section we compare the semiclassical result for the
Green function with the exact quantum result. The case $E>0$ will be
treated in App.~\ref{sec:continuum}.

\begin{figure}[t]
\begin{center}
\includegraphics[width=0.9\columnwidth]{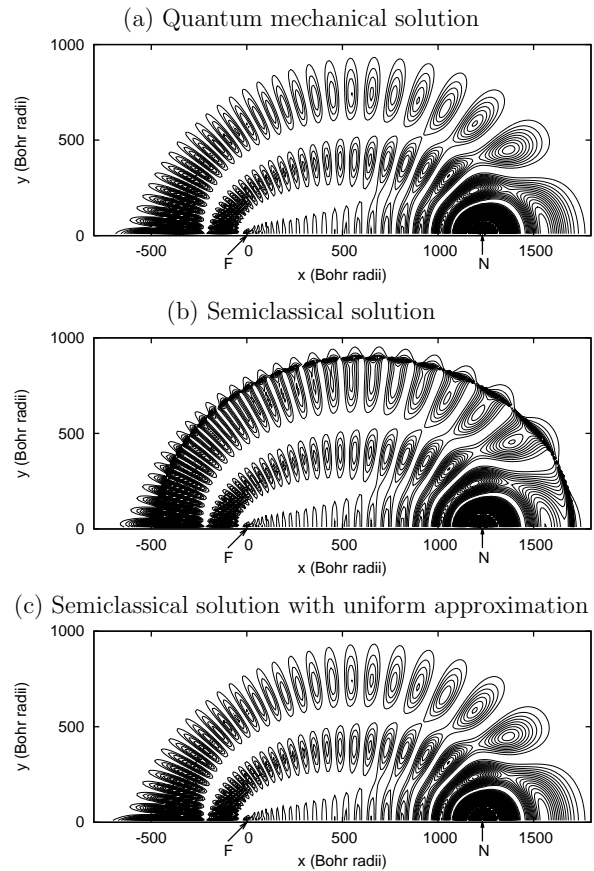}
\end{center}
\caption{Contour plot of $|G^{(3)}(\mathbf{r},\mathbf{r}';E)|^2$ with
'principal' quantum number $\nu=29.2$ and center of force $F$ at the
origin. The initial position vector $N$ at $\mathbf{r}' = (1232,0,0)$
Bohr radii is located on the $x$ axis, the final position vector
$\mathbf{r}$ is varied in the $x,y$-plane.
}\label{fig:n30x1232}\end{figure}

\section{Results and Comparison with the Exact Propagator}\label{sec:result}

\begin{figure}[t]
\centering
\includegraphics[width=0.9\columnwidth]{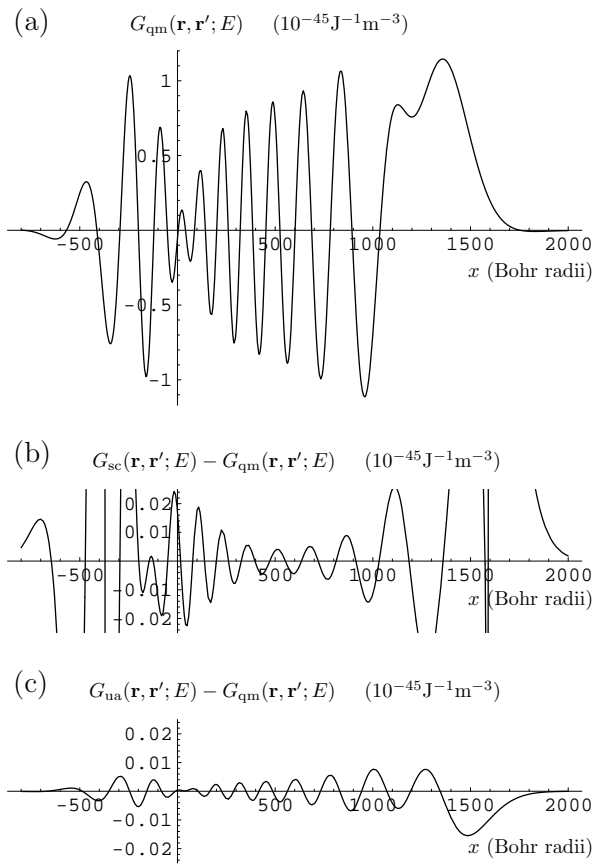}
\caption{
Plot of $G^{(3)}(\mathbf{r},\mathbf{r'};E)$ with principal
quantum number $\nu=29.2$ and center of force at the origin. The
initial position vector is located at $\mathbf{r}'=(1232,0,0)$ Bohr
radii, the final position vector shown along the $x$-axis for fixed
$y=400$ Bohr radii. (a) Quantum mechanical Green function,
(b) deviation of the semiclassical Green function from the quantum mechanical one,
(c) deviation of the uniformly approximated Green function from the quantum mechanical result.
The uniform approximation agrees with the quantum mechanical result better than $1/100$.}
\label{fig:2dschnittellipse}
\end{figure}

In a last step we compare the analytic expressions for the Coulomb
Green function with the exact quantum mechanical Green function as a
function of $\mathbf{r}$. We use atomic units.
Fig.~\ref{fig:n30x1232} shows contour plots of the three dimensional
Green function $G(\mathbf{r},\mathbf{r}';E)$. In order to avoid the infinities
at integer principal quantum numbers, we have chosen the
non--integer 'principal' quantum number $\nu = k + 1 = 29.2$ in
Eq.~(\ref{eq:energyvalues}). This value is close to the one treated
numerically in \cite{Granger2001a}.  The center of force is located
at the origin, the starting point with $\mathbf{r'} =(1232,0,0)$
Bohr radii has been chosen to lie on the $x$ axis. The end point
$\mathbf{r}$ is varied in the $x,y$- plane. \\
To illustrate the meaning of $G$ we assume to have a
coherent stationary source $\sigma(\mathbf{r}')$ of independent
particles. Such a source will generate the following wave function:
\begin{equation}
\psi(\mathbf{r},E)=\int G(\mathbf{r},\mathbf{r}';E) \, \sigma(\mathbf{r}')\; \rmd\mathbf{r}'.
\end{equation}
For a point like source at $\mathbf{r}'$ the plot of the Green function
reveals how particles leak out of the point source at $ \mathbf{r'}$
under the influence of the Coulomb field. In our case $G$ is real;
hence there is no net current flowing out of the source. All
particles are eventually reflected back into the source. A
comparison with the exact Green function shows that all features,
including the nodal structure are mirrored perfectly by the
semiclassical Green function. However we must face the fact that the
semiclassical approach will fail at the caustic where two
trajectories merge into one. Here the deficiency  can be repaired by
making use of the uniform approximation~(see App.~\ref{sec:quantities}). 
The uniform approximation is slightly more
complex than the semiclassical approximation.\\
To demonstrate how well the approximation works we present
a cut of the Green function parallel to the $x$--axis
(Fig.~\ref{fig:2dschnittellipse}). The semiclassical approximation
starts to deviate from the exact solution near the caustic where the
saddle point approximation that underlies the semiclassical theory
is no longer valid. The spike in the figure marks the position of
the caustic. At the caustic the semiclassical approximation should
be replaced by the uniform approximation which is seen to match the
exact quantum solution very well.\\
The mapping of the Coulomb problem to a 1-D--problem has
the great advantage that \textit{tunneling} properties in a Coulomb
field can be easily calculated in semiclassical approximation
because one can avoid the inherent difficulties associated with
multidimensional tunneling. Tunneling trajectories are shown in
Fig.~\ref{fig:ZweiPfade}. In the tunneling region there is exponential decay but no
reflection. The analytic continuation of the action into the
tunneling sector is given in Appendix \ref{sec:quantities}. The same
projection method as before can be used. This time the Morse
indices are no longer integers and will depend on how deep the
particle will move into the tunnel.

\section{Concluding Remarks}

Lambert's theorem has proven powerful for calculating the
semiclassical Green function (Eq.~(\ref{eq:Greencomplete})) because
it allows to parameterize all dynamical quantities in terms of
distances. This feature allowed us to eliminate the eccentric
anomaly which is ambiguous in the angles and therefore has to be
treated very carefully~\cite{Plummer1909a}. The $n$-dimensional
Coulomb problem could be reduced to one dimensional motion. The
reduction is achieved by the introduction of new variables
$\alpha_{\pm}=r+r'\pm s$. All necessary reduced actions could then
be found analytically. We derived a closed expression for the
semiclassical Green function. The  Morse indices followed directly
from the analytic form of the Van Vleck--Pauli--Morette determinant.\\
The semiclassical energy Green function is found to be an
excellent approximation to the exact Green function. It also yields
the correct bound--state energies for hydrogen in all dimensions. We
should point out that the semiclassical approximation also works
very reliably even at low energies with small principal quantum
numbers. We found that also in those cases semiclassics matches the
quantum mechanical Green function extremely well. For small quantum
numbers one has less nodes and the elliptically shaped caustic
shrinks.\\
In energetically forbidden regions there are no classical
trajectories. Nevertheless we can continue the semiclassical Green
function into the tunnel. The exit of the tunnel can be dealt with
in semiclassics by invoking corrections given by  the uniform
approximation.\\
The motion in a Coulomb potential is an important problem
in quantum mechanics. It is therefore useful to learn how the
semiclassical limit of the energy Coulomb Green function emerges
from a coherent summation of all amplitudes that belong to an
infinite number of classical trajectories. The results of this paper
can be readily implemented into real--space problems in the presence
of Coulomb interaction. One example is the quantum behavior of
Rydberg atoms \cite{Granger2001a,Ates2008a}.

\begin{acknowledgements}
This work was supported by DFG grant KL 315/7--1 and the Emmy-Noether program of the DFG (grant KR 2889/2-1).
We appreciate helpful discussions with Eric J.\ Heller, Erich Mueller, and Jan M.\ Rost.
\end{acknowledgements}

\appendix

\section{Reduced Coulomb Action}\label{sec:transformation}

In order to eliminate the eccentricity $\epsilon$ in
Eq.~(\ref{eq:action1}) in favor of spatial positions we introduce in
a first step the new variables
\begin{equation}
\cos g := \epsilon \cos\left(\frac{\xi + \xi'}{2}\right) \quad {\rm and} \quad
d := \frac{\xi - \xi'}{2} \,.
\end{equation}
In a second step we substitute
\begin{equation}
\gamma:= d+g \quad {\rm and}  \quad \delta := g - d
\end{equation}
to arrive at Eq.~(\ref{eq:action2}). We next relate the variables 
$\gamma$ and $\delta$ to spatial positions. The radial position of
any point $M$ or $N$ on the ellipse (see Fig.~\ref{fig:lambert})
relative to the center of force is given by
\begin{equation}
r^2 = [(x-\epsilon a)^2 + y^2],
\end{equation}
where $x$ and $y$ are the coordinates relative to the center of the
ellipse. With the help of Eq.~(\ref{eq:Cartesian}) we easily find
\begin{equation}
r = a(1-\epsilon \cos\xi)
\end{equation}
In terms of the variables $g$ and $d$ we have
\begin{equation}
r + r' = 2a(1- \cos g \cos d)
\end{equation}
and
\begin{equation}
|\mathbf{r} - \mathbf{r^{\prime}}| = 2a |\sin d\cos g|
\end{equation}
Without loss of generality we can assume $0\leq d \leq \pi$ and
$0\leq g \leq \pi/2$. From the last two equations we then readily
confirm the desired result, Eq.~(\ref{eq:definition}).

\section{$E>0:$ Scattering States}\label{sec:continuum}

To treat scattering states in semiclassical approximation we can use
the same formalism as for bound states. In an attractive force field
and for $E>0$, there is no caustic and hence no reflection at $b$.
As can be seen from Fig.~\ref{fig:Laufwege3} we then have only two
hyperbolic trajectories leading from $N$ to $M$. The one--dimensional variables are again
$\alpha_\pm=r+r'\pm|\mathbf{r}-\mathbf{r}'|$. The projection of the motion
onto a line applies again but we have to consider the change in
geometry.

\subsection{Attractive Coulomb Interaction}

In this case we obtain
\begin{widetext}
\begin{equation}\label{eq:anzhyperbel}
\begin{split}
G_{\rm{sc,attr}}^{(n)}(\mathbf{r},\mathbf{r}';E)& = -\frac{\rmi}{\hbar}
\frac{1}{(2\pi\rmi\hbar)^{(n-1)/2}}\\
&\times \left(\sqrt{|\mathbf{D}\encircle{1}|}
\exp[\frac{\rmi}{\hbar}W_1]+\sqrt{|\mathbf{D}\encircle{2}
|}\exp[\frac{\rmi}{\hbar}W_2-\rmi\frac{\pi}{2}(n-2)]\right)\,,
\end{split}
\end{equation}
\begin{figure}[t]
\begin{center}
\includegraphics[width=0.22\textwidth]{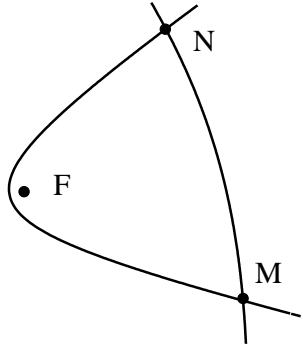}
\end{center}
\caption{For $E>0$, hyperbolic trajectories connect $N$ and $M$, with the center of force at $F$.
In the repulsive case (not shown) there is a caustic in contrast to the attractive case where every point in space
can be reached.}
\label{fig:Laufwege3}
\end{figure}
with $W_1=W_+-W_-$ and $W_2=W_++W_-$. The action follows again from
Eq.~(\ref{eq:wirkungen1}), adapted to $E>0$,
\begin{equation}\label{eq:wirkungen3}
\begin{split}
\left(W_\pm\right)_{\rm{attr}}^{(E>0)}&=\mu\sqrt{\frac{K_c}{\mu
a}}\int_{2 a}^{\alpha_\pm/2}
    \sqrt{\frac{2 a +\tilde{\alpha}_\pm/2}{\tilde{\alpha}_\pm/2}}\,\rmd (\tilde{\alpha}_\pm/2)\\
    &=\sqrt{\frac{K_c \mu}{a}}\left(\frac{1}{2}\sqrt{(4a+\alpha_\pm)\alpha_\pm}+ 2a
    \log{\frac{\sqrt{\alpha_\pm}+\sqrt{4a+\alpha_\pm}}{\sqrt{4a}}}\right).
\end{split}
\end{equation}
\end{widetext}

\subsection{Repulsive Coulomb Interaction}

If the potential is repelling we have again two hyperbolic
trajectories which connect $N$ and $M$. But now a caustic separates
the classically allowed region from the energetically forbidden
region. Classically allowed motion occurs for $4 |a|<\alpha_\pm$.
The corresponding reduced action reads
\begin{widetext}
\begin{equation}\label{eq:wirkungen4}
\begin{split}
\left(W_\pm\right)_{\rm{rep}}^{(E>0)}&=\mu\sqrt{\frac{K_c}{\mu
|a|}}\int_{2 a}^{\alpha_\pm/2}
    \sqrt{\frac{-2 |a| +\tilde{\alpha}_\pm/2}{\tilde{\alpha}_\pm/2}}\,\rmd (\tilde{\alpha}_\pm/2)\\
    &=\sqrt{\frac{K_c \mu}{|a|}}\left(\frac{1}{2}\sqrt{(-4 |a|+\alpha_\pm)\alpha_\pm}+|a|
    \log{\frac{\sqrt{\alpha_\pm}-\sqrt{-4 |a|+\alpha_\pm}}{\sqrt{\alpha_\pm}+\sqrt{-4
    |a|+\alpha_\pm}}}\right).
\end{split}
\end{equation}
For completeness we write down the action in the classically
forbidden tunneling region where $\alpha_- < 4|a|$:
\begin{equation}\label{eq:wirkungen5}
\begin{split}
\left(W_-\right)_{\rm{rep}}^{(E>0)}&=\pm\rmi\mu\sqrt{\frac{K_c}{\mu
|a|}}\int_{2 a}^{\alpha_-/2}
    \sqrt{\frac{2 |a| -\tilde{\alpha}_-/2}{\tilde{\alpha}_-/2}}\,\rmd (\tilde{\alpha}_-/2)\\
    &=\pm\rmi\sqrt{\frac{K_c \mu}{|a|}}\left(-\pi|a|+
    \frac{1}{2}\sqrt{(4 |a|-\alpha_-)\alpha_-}+ 2
    |a|\arctan{\sqrt{\frac{\alpha_-}{4|a|-\alpha_-}}}\right).
\end{split}
\end{equation}
\end{widetext}

\section{Analytic Continuation into the Tunneling Region}\label{sec:quantities}

\subsection{Uniform Approximation}

The semiclassical Green function is derived from the exact
expression for the quantum mechanical Green function by making use of the saddle
point approximation (SPA). However this approximation is not valid
at the caustic where two saddle points merge into one. In this case
the uniform approximation (UA) will cure the deficiency of the SPA.
The method is standard. For more details the reader is referred to
\cite{Schulman1981a}, p.~118ff, p.~131ff. Here we follow the method
outlined in reference \cite{Richards2002a}. For $n=3$ we have

\begin{equation}\label{eq:GUAtoll}
G_{\text{ua}}(\mathbf{r},\mathbf{r}';E) = \frac{\rme^{\rmi\xi}}{2\hbar^2\sqrt{\pi}}
\left(d_0\Ai(-\zeta)-\rmi d_1\Ai'(-\zeta)\right)
\end{equation}
with
\begin{equation}
\xi=
\begin{cases}
\frac{1}{2\hbar}\left(W_++W_-\right) & \text{if $\Im(W_-)=0$,}\\
\Re(W_+/\hbar)                       & \text{if
$\Im(W_-)=-\Im(W_+)$},
\end{cases}
\end{equation}
\begin{equation}
\zeta=
\begin{cases}
{\left[\frac{3}{4\hbar}\left(W_+-W_-\right)\right]}^{2/3} & \text{if $\Im(W_-)=0$,}\\
-\frac{3}{2}{\left[\Re(W_+/\hbar)\right]}^{2/3}           & \text{if
$\Im(W_-)=-\Im(W_+)$},
\end{cases}
\end{equation}
and
\begin{equation}
\begin{split}
d_0&=\zeta^{1/4} \left[\sqrt{\mathbf{D}(W_+)}+\sqrt{-\mathbf{D}(W_-)}\right]\rme^{-5\rmi\pi/4}\\
d_1&=\zeta^{-1/4}\left[\sqrt{\mathbf{D}(W_+)}-\sqrt{-\mathbf{D}(W_-)}\right]\rme^{-5\rmi\pi/4}.
\end{split}
\end{equation}

\begin{figure}[b]
\begin{center}
\includegraphics[width=0.7\columnwidth]{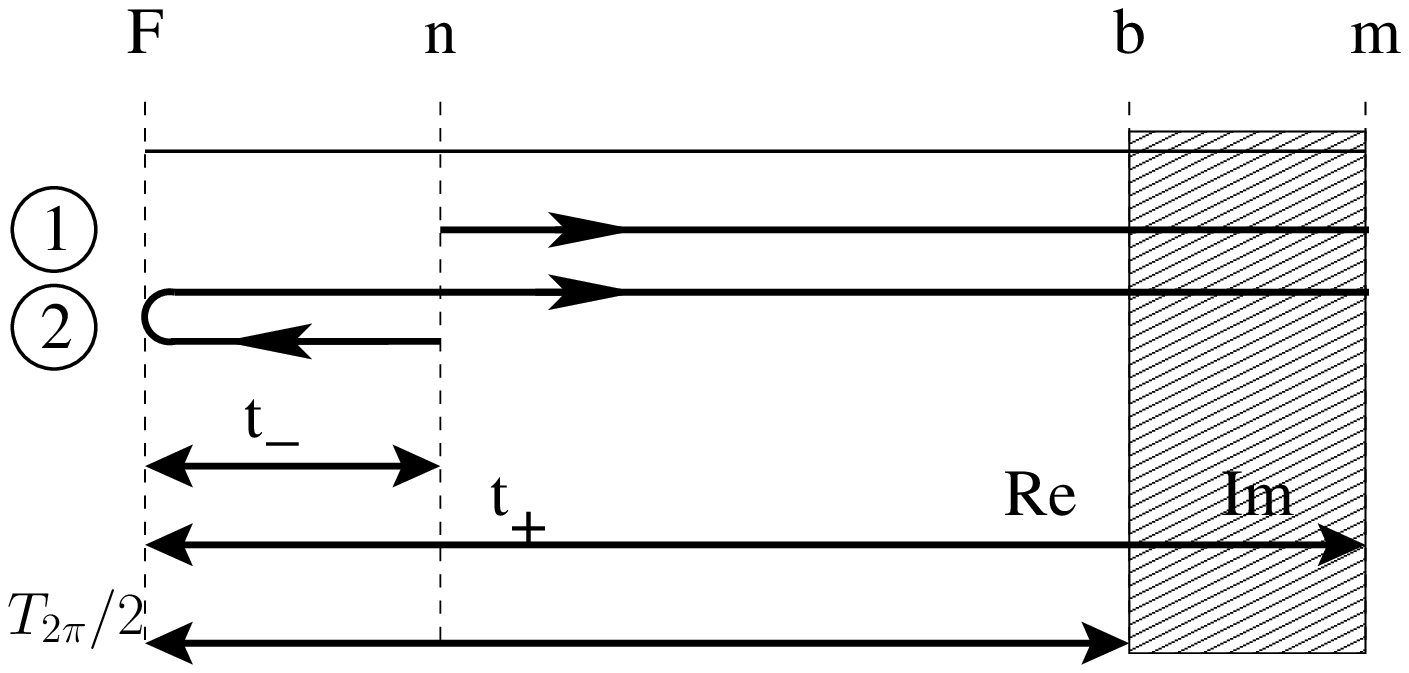}
\end{center}
\caption[Solutions for nonclassical paths]{Bound states and
tunneling trajectories: We use the projection formalism to find two
possible paths to a point in the classically forbidden region. As
the electron passes through b into the tunneling space sector,
action and velocity become complex (see Eq.~(\ref{eq:wirkungen2})).}
\label{fig:ZweiPfade}
\end{figure}

\subsection{Tunneling Regime}

Here we look at classically forbidden motion $\alpha_+>4 a$. This
means that point $m$ is lying in the tunnel (see
Fig.~\ref{fig:ZweiPfade}). Whereas $W_-$ is not changed, $W_+$ must
be continued into the classically forbidden space sector. The
analytic continuation is obtained by making use of
Eq.~(\ref{eq:wirkungen1})
\begin{widetext}
\begin{equation}\label{eq:wirkungen2}
\begin{split}
\left(W_+\right)_{\rm{forb}}&=\mu\sqrt{\frac{K_c}{\mu
a}}\left(\int_0^{2a}
    \sqrt{\frac{2 a -\tilde{\alpha}_\pm/2}{\tilde{\alpha}_\pm/2}}\,\rmd (\tilde{\alpha}_\pm/2) \pm\rmi\int_{2 a}^{\alpha_\pm/2}
\sqrt{\frac{-2 a +\tilde{\alpha}_\pm/2}{\tilde{\alpha}_\pm/2}}\,\rmd (\tilde{\alpha}_\pm/2)\right)  \\
    &=\sqrt{\frac{K_c \mu}{a}}\left(a\pi \pm\rmi\left(\frac{1}{2}\sqrt{(-4 a+\alpha_\pm)\alpha_\pm}+ 2 a \log{\frac{\sqrt{4 a}}{\sqrt{\alpha_\pm}+\sqrt{-4 a+\alpha_\pm}}}\right)\right)
\end{split}
\end{equation}
\end{widetext}
There are two complex conjugated solutions. For the propagator we
select the term with positive imaginary part to ensure that the wave
function decays exponentially deep in the tunnel. Note however that
both solutions will contribute to the uniform approximation in the
vicinity of the tunnel exit.

\providecommand{\url}[1]{#1}

\end{document}